\documentclass[epj]{webofc}
\def\xb{\overline{x}}

\begin{document}
\title{Amplitudes ratios in $\rho^0$ leptoproductions and GPDs}
%
%

\author{\firstname{S.V. Goloskokov}\inst{1}\fnsep\thanks{\email{goloskkv@theor.jinr.ru}}
}

\institute{Bogoliubov Laboratory of Theoretical Physics, Joint
Institute for Nuclear Research, Dubna 141980, Moscow region,
Russia}

\abstract{%
We investigate exclusive leptoproduction of $\rho^0$  meson. These
reactions were analyzed within the factorizing handbag approach.
In our model   good agreement of observables for light meson
production  with experimental data in a wide energy range was
found.

Using the model results we  calculate the ratio of different
helicity amplitudes for a transversely polarized proton target to
the leading twist longitudinal amplitude. Our results are close to
the amplitude ratios measured by HERMES.}
\maketitle
\section{Introduction}
\label{intro} In this report, investigation of $\rho^0$ meson
leptoproduction is based on the handbag approach. Here the
amplitudes at high $Q^2$ factorize into hard meson
electroproduction off partons and the Generalized Parton
Distributions (GPDs) \cite{fact}.  Good agreement of our results
for the cross sections and spin observables,  expressed in terms
of GPDs $H$, $E$,  with experimental data were obtained in
\cite{gk06,gk08}. We consider transversity effects $H_T$, $\bar
E_T$, that have a twist-3 character \cite{gk11}. This gives us a
possibility to describe spin observables that are equal to zero
without twist-3 contributions. Essential contributions from
unnatural parity exchanges were found by HERMES \cite{omega14} in
the $\omega$ production. It was shown  that the pion pole (PP)
contributions \cite{gk14} are significant for explanation of the
large unnatural-parity effects at HERMES in $\omega$ and $\rho^0$
production. We discuss the amplitude properties and physical
observables in section 2.

In section 3, using the model results \cite{gk06,gk08,gk11,gk14},
we  calculate the ratio of different helicity amplitudes to the
leading twist longitudinal amplitude and compare the results with
preliminary HERMES data \cite{man}.
\section{$\rho^0$ meson leptoproduction. Physical observables}
We can define  Natural and Unnatural parity NP (UP) amplitudes as
\begin{equation}\label{mn}
T(U)_{\lambda \nu',\mu \nu}=\frac{1}{2}\,\large [ M_{\lambda
\nu',\mu \nu}\pm (-1)^{\mu-\lambda}\, M_{-\lambda \nu',-\mu \nu}
\large ].
\end{equation}
Here $\nu, \nu'$ are the initial and final proton helicities and
$\mu, \lambda$ are the helicities of the photon and final $\rho^0$
meson. In what follows, according to \cite{herm1}, we  use the
initialisms for the  proton spin-non flip and  spin-flip
amplitudes, respectively,
\begin{equation}\label{nu}
T(U)^{(1)}_{\lambda,\mu}=N(U)_{\lambda + ,\mu + };\;\;\;
T(U)^{(2)}_{\lambda,\mu}=N(U)_{\lambda - ,\mu + }.
\end{equation}

 We calculate $\rho^0$ leptoproduction off
proton within the handbag approach where the leading  amplitude
at  high photon virtuality $Q^2$ can be represented in a
factorized form \cite{fact} as a convolution of a hard meson
subprocess amplitude off partons ${\cal H}^a_{\mu' +,\mu +}$,
which is calculated perturbatively, and GPDs as
\begin{equation}\label{ff}
T^{(1)}_{\lambda \mu } \propto \int_{-1}^1 d\xb\,
   {\cal H}^a_{\lambda +,\mu +} F^a(\xb,\xi,t);\;\; T^{(2)}_{\lambda \mu } \propto
   -\frac{\sqrt{-t'}}{2 m}\int_{-1}^1 d\xb\,{\cal H}^a_{\lambda +,\mu
                          +}\,
           E^a(\xb,\xi,t).
\end{equation}
Here $a$ is a flavor factor. Generally, factorization is  not
valid for the not leading amplitudes. These problems can be solved
in our model \cite{gk06,gk08} where subprocesses are calculated
within the modified perturbative approach \cite{sterman} in which
quark transverse degrees of freedom accompanied by Sudakov
suppressions are considered. The quark transverse momentum
regularizes the end-point singularities in the TT amplitude thus
it can be calculated.

GPDs  contain  broad information on the hadron structure. With the
help of sum rules GPDs are connected with the hadron form factors,
and information on the parton angular momenta can be extracted. In
the forward limit $t=0$ and zero skewness $\xi=0$ GPDs are
equivalent to ordinary Parton Distribution Functions (PDFs). The
GPDs are estimated by us using the double distribution
representation \cite{mus99}, which connects  GPDs with PDFs
through the double distribution function $\omega$. For the valence
quark contribution  $\omega$ looks like
\begin{equation}\label{ddf}
\omega_i(x,y,t)= h_i(x,t)\, \frac{3}{4}\, \frac{[(1-|x|)^2-y^2]}
                           {(1-|x|)^{3}}.
\end{equation}
 The functions $h$ in (\ref{ddf}) are
parameterized as PDFs   in the form
\begin{equation}\label{pdfpar}
h(x,t)= N\,e^{b_0 t} x^{-\alpha(t)}\,(1-x)^{n},
\end{equation}
with the $t$- dependence which is considered in a Regge form, and
$\alpha(t)$ is the corresponding Regge trajectory. The parameters
 for PDFs $h$ in (\ref{pdfpar}) are obtained from the analyses
\cite{CTEQ6}; information about PDFs $e$ is taken from
\cite{pauli}. The handbag approach describes  successfully the
light meson leptoproduction at HERMES, COMPASS and HERA energies
\cite{gk06,gk08}. It can be seen from Fig. 1, (left) that we
describe properly the energy dependence of the $\rho^0$ cross
section from HERMES to HERA energies. Here GPDs $H$ are essential.
At $W < 5 \mbox{GeV}$ the cross section grows unexpectedly. This
effect is not understood.

In Fig. 1, (right) we show the $A_{UT}^{\sin(\phi-\phi_s)}$
asymmetry for $\rho^0$ production at COMPASS energy which is
determined mainly by the interference of GPDs $H$, $E$
\begin{equation}\label{w}
A_{UT}^{\sin(\phi-\phi_s)} \sim  \mbox{Im}\large [ <E>^*\, <H>
\large ].
\end{equation}
We describe properly $t$ dependence of
$A_{UT}^{\sin(\phi-\phi_s)}$ asymmetry at COMPASS.

\begin{figure}[h]
\centering
\begin{tabular}{cc}
\includegraphics[width=6.4cm,height=4.8cm]{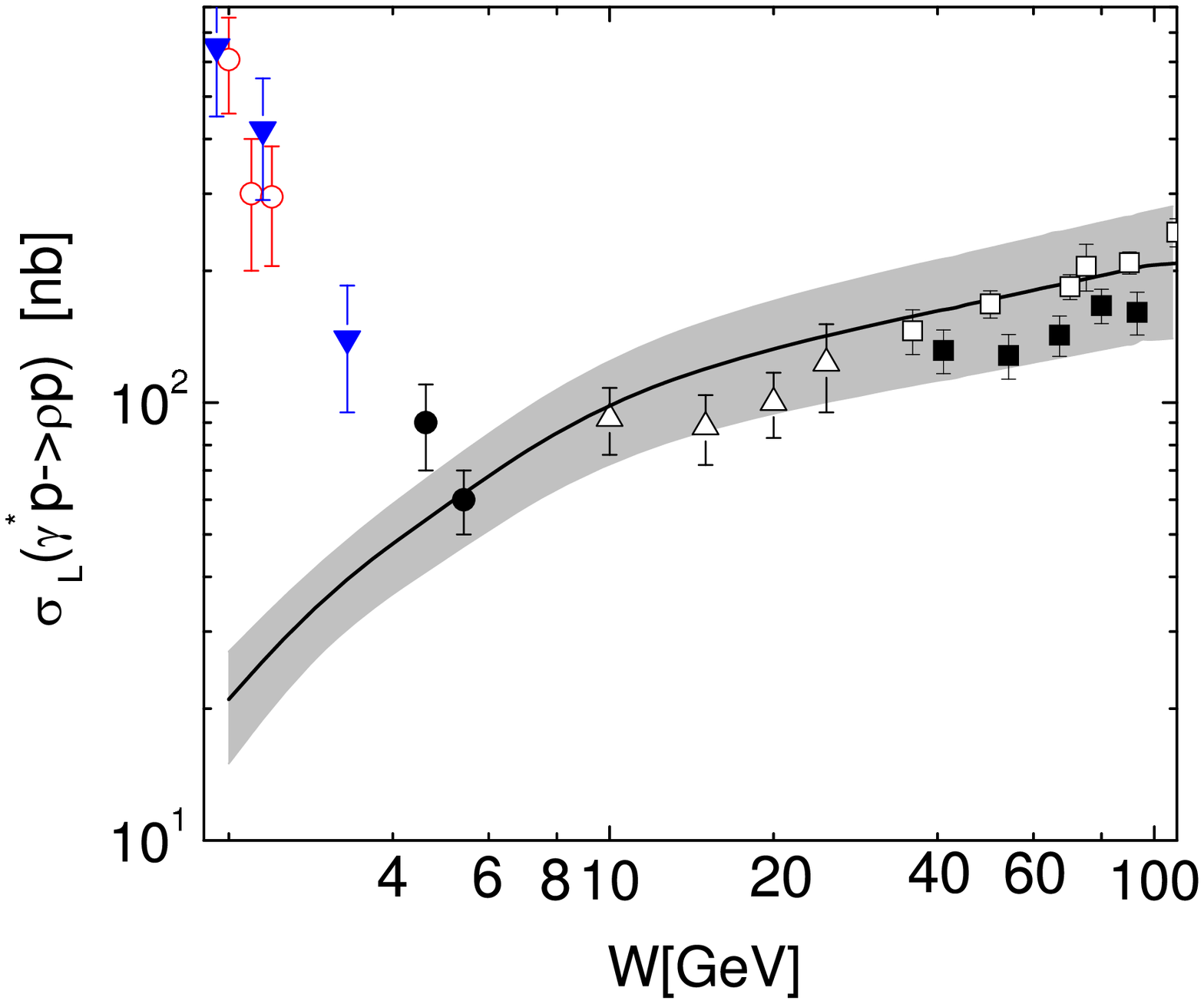}&
\includegraphics[width=6.7cm,height=5.1cm]{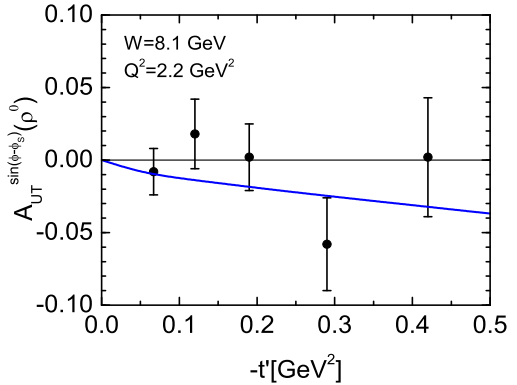}
\end{tabular}
\caption{{\bf Left}: Longitudinal $\rho^0$ cross section at
$Q^2=4.0\,\mbox{GeV}^2$. HERMES (solid circle), ZEUS (open
square), H1 (solid square), E665 (open triangle), open circles-
CLAS,  CORNEL -solid triangle. {\bf Right}: Model results for
$A_{UT}^{\sin(\phi-\phi_s)}$ asymmetry for $\rho^0$ production
with COMPASS data \cite{compaut}.}
\label{fig-1}       
\end{figure}

\begin{figure}[h]
\centering
\begin{tabular}{cc}
\includegraphics[width=6.7cm,height=5cm]{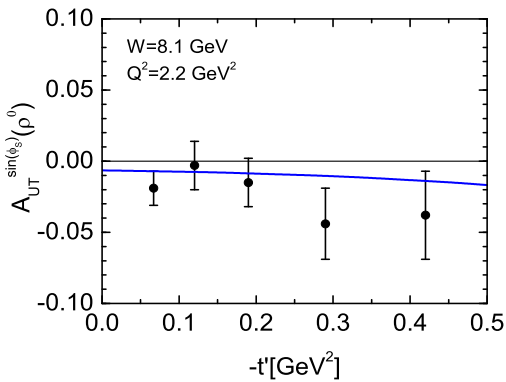}&
\includegraphics[width=6.7cm,height=5cm]{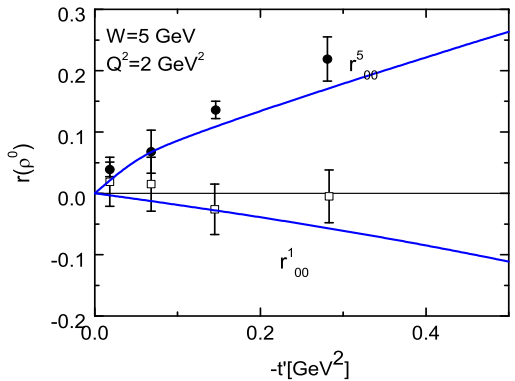}
\end{tabular}
\caption{{\bf Left}: Model results for $A_{UT}^{\sin(\phi_s)}$
asymmetry of $\rho^0$ production with COMPASS data \cite{compaut}.
{\bf Right}: $\bar E_T$ effects in SDMEs of $\rho^0$ production.
HERMES data are shown \cite{hermsdme}.}
\label{fig-2}       
\end{figure}

Unfortunately, some asymmetries and Spin Density Matrix Elements
(SPMEs), which  are not small experimentally, are equal to zero in
the leading-twist approximation of our GPD model. To describe the
experimental data on the meson electroproduction  at low $Q^2$, we
consider the amplitudes $T^{(1,2)}_{0 1}$, which are determined in
terms of the transversity GPDs $H_T$ and $\bar E_T$. Within the
handbag approach the transversity GPDs are accompanied by a
twist-3 meson wave function in the hard subprocess amplitude
${\cal H}$ \cite{gk11}, which is the same for both the
$M^{(1,2)}_{0 1}$ amplitudes:

\begin{equation}\label{ht}
  T^{(1)}_{0 1} \propto \, \frac{\sqrt{-t'}}{4 m}\,
                            \int_{-1}^1 d\xb
 {\cal H}_{0-,++}\; \bar E_T(\xb,\xi,t);\;M^{(2)}_{0 1} \propto \,
                            \int_{-1}^1 d\xb
   {\cal H}_{0-,++}\,H_T(\xb,\xi,t).
\end{equation}

The $H_T$ GPDs in the forward limit and $\xi=0$ are equal to
transversity PDFs $\delta$ and are parameterized   by using the
model \cite{ans}. Information on $\bar E_T$ is obtained now only
from the lattice QCD \cite{lat}. We estimate the corresponding
$\bar e_T$ PDFs from the lattice results using form
(\ref{pdfpar}). The double distribution  is used to calculate
transversity GPDs as before. Note that the amplitude $M^{(2)}_{0
1}$ has no definite parity. Really, $M^{(2)}_{0
1}=M_{0-++}=-T^{(2)}_{0 1}+U^{(2)}_{0 1}$ and
$M_{0--+}=-T^{(2)}_{0 1}-U^{(2)}_{0 1}$. The last amplitude is
equal to zero in the model and $T^{(2)}_{0 1}=-U^{(2)}_{0 1}$.
Thus  NP and UP contributions to the $M^{(2)}_{0 1}$ amplitude
have the same value.

In Fig. 2, we show transversity effects in spin observables. In
Fig. 2, (left) the $A_{UT}^{\sin(\phi_s)}$ asymmetry is presented
together with COMPASS data. This asymmetry is determined by the
$H$, $H_T$ interference
\begin{equation}\label{q}
A_{UT}^{\sin(\phi_s)} \sim \mbox{Im}[<H_T>^* <H>].
\end{equation}
The data are described quite well by the model results. In Fig. 2,
(right) the SDMEs $r^5_{00}$ and $r^1_{00}$ are shown. The first
SDME is determined by the $H$, $\bar E_T$ interference and the
second one by the $\bar E_T$ contribution only
\begin{equation}\label{1}
r^5_{00} \sim \mbox{Re}[<H>^* <\bar E_T>];\;\;\; r^1_{00} \sim
-<\bar E_T>|^2.
\end{equation}
We find that we describe properly both SDMEs at HERMES.

Now we shall discuss UP effects. The UP contribution to the
$U^{(1)}_{1 1}$ amplitude  is determined by the polarized $\tilde
H$ quarks GPDs and can be observed in the $A_{LL}$ asymmetry
\cite{gk08}. It was found that the $\tilde H$ effects provide the
$A_{LL}$ asymmetry that is a little bit smaller with respect to
experiment \cite{gk08}.

The HERMES data for the $\omega$ production indicate the strong
contributions from UP effects \cite{omega14}. This can be seen,
e.g., from the ratio of the unnatural and natural parity cross
sections, which was found to be larger than unity. This effect can
be caused \cite{gk14} by the large PP contribution to this
process.

 The UP helicity amplitudes determined by PP contribution to
$\rho^0$ leptoproduction looks as follows \cite{gk14}. The
dominant contributions are:
\begin{equation}\label{uu}
U^{(1)}_{1 1} \sim \frac{\rho_{\pi \rho}}{t-m^2_\pi}\,
\frac{m\,\xi\, Q^2}{\sqrt{1-\xi^2}},\;\; U^{(2)}_{1 1} \sim
-\frac{\rho_{\pi \rho}}{t-m^2_\pi}\,\frac{\sqrt{-t'}\,Q^2}{2}.
\end{equation}
Here $\rho_{\pi \rho}$ is proportional to the $\pi \rho$
transition form factor $g_{\pi \rho}$. It  can be determined from
the $\rho^0$ meson radiative decay. For $\rho^0$ production the
absolute value of the $\pi \rho$ transition form factor (FF) at
zero photon virtuality can be estimated as \cite{gk14}
\begin{equation}
\Gamma(\rho \to \pi \gamma) \sim \frac{\alpha_{elm}}{24}\,|g_{\pi
\rho}(0)|^2 M_V^3;\;\; |g_{\pi \rho}(0)|=.85\mbox{GeV}^{-1}.
\end{equation}
This value of $g_{\pi \rho}$ was used in \cite{gk14} in analyses
of PP effects in $\rho^0$ production.

\section{Amplitude ratios of $\rho^0$ meson leptoproduction.}

\begin{figure}[h]
\centering
\includegraphics[width=13.5cm,height=10cm]{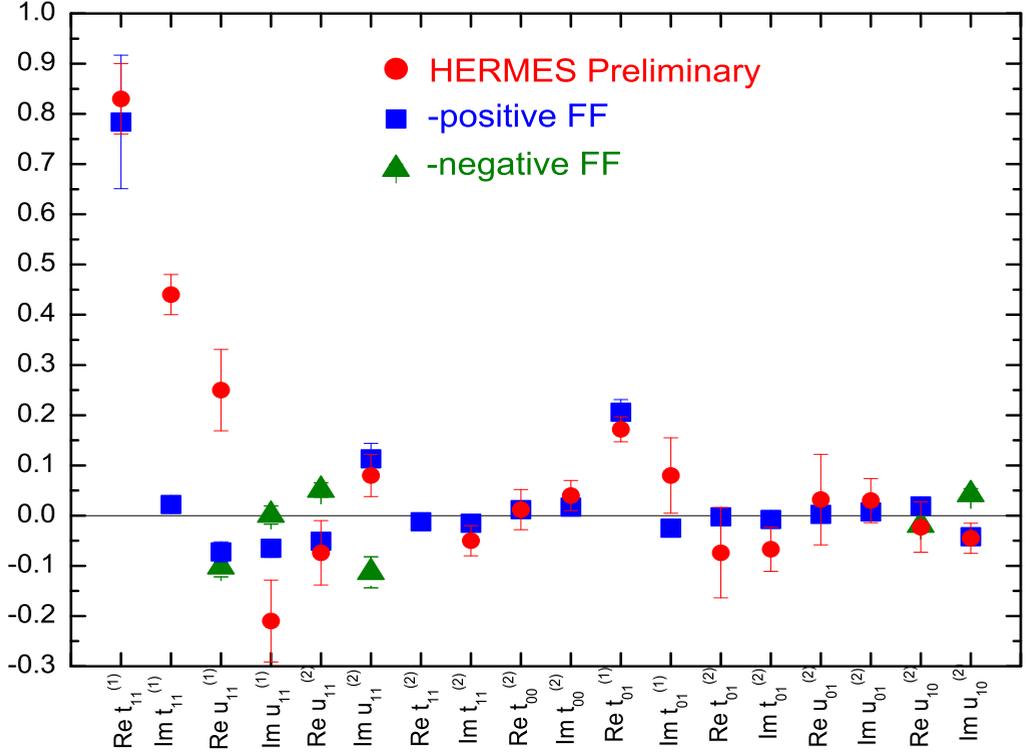}
\caption{Model results for amplitude ratios of $\rho^0$ meson
leptoproduction for positive and negative $\pi \rho$ transition FF
together with HERMES preliminary data \cite{man}.}
\label{fig-3}       
\end{figure}

Our GPD model with NP amplitudes determined by $H,\; E,\; H_T,\;
\bar E_T$ GPDs and UP effects caused by  $\tilde H$ and PP
contributions describes well physical observables
\cite{gk06,gk08,gk11,gk14}. However, observables  in terms of GPDs
have usually a quite complicated form. More direct information
about GPDs can be found from the ratio of different helicity
amplitudes on transversely polarized proton target to the leading
twist longitudinal amplitude

\begin{equation}\label{rt}
t^{(i)}_{\lambda \mu}=T^{(i)}_{\lambda \mu}/T_{00},\;\;\;
u^{(i)}_{\lambda \mu}=U^{(i)}_{\lambda \mu}/T_{00},
\end{equation}
which can be calculated using the model results
\cite{gk06,gk08,gk11,gk14}.

To be consistent with the HERMES analyses, we study the  ratios
$t(u)^{(i)}_{\lambda ,\mu }$ in the handbag approach using the
HERMES definitions of the amplitude signs. The UP amplitudes
(\ref{uu}) are dependent on the $g_{\pi \rho}$ sign. The model
results for positive $g_{\pi \rho}$ transition FF (squares) and
negative FF (triangles) together with HERMES data are shown in
Fig. 3. We present here only the amplitude ratios which are not
zero in the model.

\begin{itemize}
 \item $t^{1}_{11}$ is the ratio of transverse to longitudinal amplitudes dominated
by $H$ GPDs. The amplitudes are mainly imaginary that give large
$\mbox{Re} t^{1}_{11}$, which is consistent with data. ${\rm Im}
t^{1}_{11}$ is rather small in the model. That is caused by the
small relative phase between transverse to longitudinal amplitudes
\cite{gk08}.
  \item Small value of $u^{1}_{11}$ can be related to  not so large value of $\tilde H$
  and the corresponding PP contribution. See the discussion of the $A_{LL}$ asymmetry.

\item  $u^{2}_{11}$ is determined by PP effects. We have found a good result for the positive
$\pi \rho$ transition FF.

\item $t^{2}_{11}$ is connected with the $E$ GPDs contribution to the proton
spin-flip amplitude for the transversely polarized photon and
meson. The result is good for the imaginary part of the ratio.
There are no experimental data for the real part.

 \item $t^{2}_{00}$- is determined by the $E$ GPDs contribution  to the longitudinal amplitude.
The model results are not far from experiment.

\item $t^{1}_{01}$ is connected  with the twist-3 transversity $\bar E_T$ effects. Our results
are consistent with HERMES data.

 \item $t^{2}_{01}$ and $u^{2}_{01}$ are determined by the twist-3 transversity $H_T$
 effects. It was mentioned that we have connection
$t^{2}_{01}=-u^{2}_{01}$ for these amplitudes. The amplitude
ratios are in agreement with the model results, but they may be a
little bit larger.

 \item  $u^{2}_{10}$- is determined by PP contribution. The obtained results are consistent with experiment
  for the positive $\pi \rho$ transition FF
\end{itemize}

We have analyzed meson electroproduction within the handbag
approach. Modified perturbative approach was used to calculate the
hard subprocess amplitude. GPDs were calculated using  PDFs on the
basis of the double distribution representation. Good description
of different spin observables was found in the model
\cite{gk06,gk08,gk11,gk14}. The ratio of different helicity
amplitudes on the transversely polarized proton target to the
leading twist longitudinal amplitude was calculated. The PP
contribution to the UP amplitudes ratios is dependent on the sign
of the $\pi \rho$ transition FF. The model results for the
positive sign of the $\pi \rho$ transition FF are compatible with
HERMES data. For the negative sign of FF the results are worse. We
can conclude that the positive sign of transition FF is
preferable. This is consistent with conclusion  done in
HERMES and COMPASS papers \cite{omega14,comp16}.\\

The work was supported in part by the Heisenberg-Landau program.

\end{document}